\documentclass[12pt]{article}
\usepackage{epsfig}
\usepackage{amsfonts}
\usepackage{amsmath}
\usepackage{amssymb}
\usepackage{bm}
\usepackage{cite}
\usepackage{rotating}
\usepackage{array}
\usepackage{multirow}
\usepackage[title]{appendix}
\begin{document}
\sffamily
\title{
Resonance states $0^+$ of the Boron isotope ${}^8\text{\sffamily B}$ 
from the Jost-matrix analysis of experimental data
}
\author{
S.A. Rakityansky$^{\,a}$\footnote{e-mail: rakitsa@up.ac.za},
S.N. Ershov$^{\,b}$\footnote{e-mail: ershov@theor.jinr.ru},
T.J. Tshipi$^{\,a}$\\[3mm]
\parbox{11cm}{%
${}^a${\small Department of Physics, University of Pretoria, Pretoria,
South Africa}\\
${}^b${\small Joint Institute for Nuclear Research, Dubna, Russia}
}
}
\maketitle
\begin{abstract}
\noindent
The available $R$-matrix parametrization of experimental data 
on the excitation functions for the elastic and inelastic $p\,{}^7\mathrm{Be}$ 
scattering
at the collision energies up to 3.4\,MeV is used to generate the 
corresponding partial-wave cross sections in the states with $J^\pi=0^+$. Thus 
obtained data are 
considered as experimental partial cross sections and are fitted using 
the semi-analytic two-channel Jost matrix with proper analytic structure and 
some adjustable parameters. Then the spectral points are sought as zeros of the 
Jost matrix determinant (which correspond to the $S$-matrix poles) at complex 
energies. The correct analytic structure makes it possible to calculate the 
fitted Jost matrix on any sheet of the Riemann surface whose topology involves 
not only the square-root but also the logarithmic branching caused by the 
Coulomb interaction. In this way, two overlapping $0^+$ resonances at the 
excitation energies $\sim1.79$\,MeV and $\sim1.96$\,MeV have been found.
\end{abstract}

\section{Introduction}
\begin{figure}
\centerline{\epsfig{file=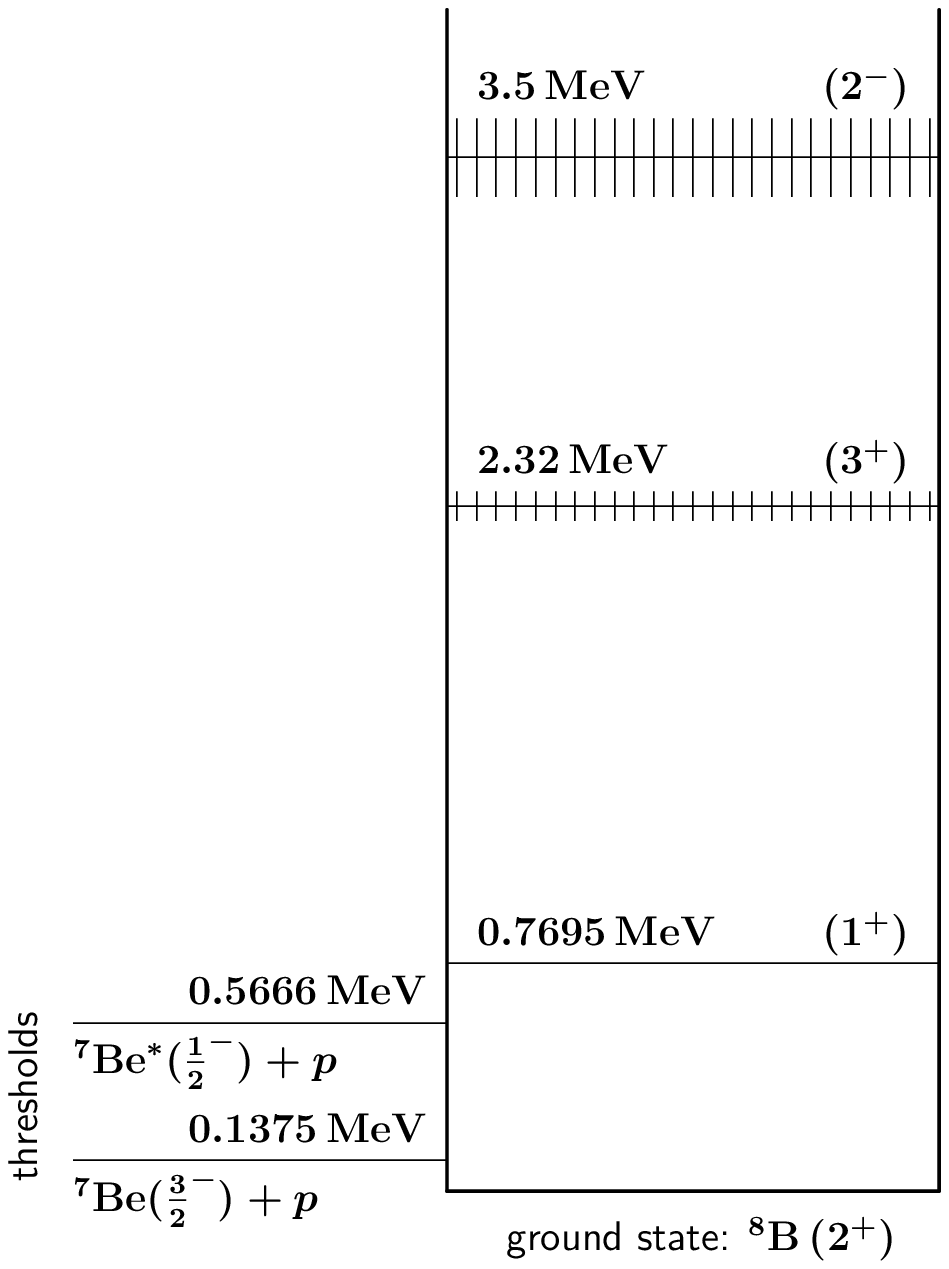}}
\caption{\sf
Low-lying excited states of the neuclear system ${}^8\mathrm{B}$ and the 
two-body thresholds for the decays 
${}^8\mathrm{B}\to{}^7\mathrm{Be}(\frac32^-)+p$ and
${}^8\mathrm{B}\to{}^7\mathrm{Be}^*(\frac12^-)+p$. The data are taken from 
Refs.~\cite{Tilley_8,Tilley_5_6_7}.
}
\label{fig.establishedlevels}
\end{figure}

The spectrum of the eight-nucleon system ${}^8\mathrm{B}$ was established as a 
result of many experimental and theoretical studies (an extensive list of 
publications related to  ${}^8\mathrm{B}$, can be found 
in Ref.~\cite{Tilley_8}). There are excited states 
that appear in all of them more or less at the same energies. These well 
established levels (taken from Ref.~\cite{Tilley_8}) are schematically depicted 
in Fig.~\ref{fig.establishedlevels}.
However, there are 
also few unconfirmed levels that follow from some of the studies and not from 
the others. In the present work, we do an attempt to clarify the existence and 
parameters of the resonance level with the quantum numbers $J^\pi=0^+$. 
Using available experimental data, we construct the two-channel Jost matrices 
and then analytically continue them onto the Riemann surface of the energy. The 
spectral points are sought as the zeros of the Jost-matrix determinant, which 
correspond to the poles of the $S$-matrix.

As the experimental data, we use the partial cross sections with $J^\pi=0^+$ 
for all four possible
transitions between the states $p\,{}^7\mathrm{Be}(\frac32^-)$ and
$p\,{}^7\mathrm{Be}^*(\frac12^-)$, where the second channel involves the first 
excited state of the Beryllium isotope. In order to obtain these 
cross sections, we use the $R$-matrix
given in Ref.~\cite{Rogachev2013}. That $R$-matrix was constructed by 
parametrizing the measured excitation functions for the elastic and inelastic 
$p\,{}^7\mathrm{Be}$ scattering.

We fit these data using
the two-channel Jost matrix. Since the cross sections are extracted from 
the available $R$-matrix parametrization of a large collection of experimental 
data, we indirectly fit the original data. After the fitting at real energies, 
the Jost matrix is considered at complex $E$, where the zeros of its determinant 
correspond to the poles of the $S$-matrix.

The milti-channel Jost matrix is taken in a special representation suggested in
Refs.~\cite{our.MultiCh, our_Coulomb}, where it is given as a sum
of two terms. Each of these term is factorized in a product of two matrices, one
of which is an unknown analytic single-valued function of the energy
and the other is given explicitly as a function of the channel momenta. The 
explicitly given factors are responsible for the branching of the Riemann 
surface. The unknown single-valued matrices are parametrized and the parameters 
are found via fitting the available experimental data.

With the semi-analytic representation of the Jost matrix, where the
factors responsible for the topology of the Riemann surface are given
explicitly, it is easy to explore the behaviour of the Jost matrix on all the
sheets of the Riemann surface. In this way we are able to accurately locate
the resonance poles and to examine the possibility that the so called
``shadow'' poles exist on the other sheets of the surface.

\section{Jost matrices}
The Jost matrices are only defined for the binary reactions, where the colliding 
particles may either change their internal states ($a+b\to a^*+b^*$) or transit
to another pair of particles ($a+b\to c+d$). In general, the masses of 
the particles may change. This means that the channels $ab$, $a^*b^*$, $cd$, 
etc. have different thresholds. 

For a given two-body system of particles, $a$ and $b$, there are infinite 
number of possible 
combinations of their orbital angular momentum $\ell$ and the two-body 
spin $\vec{s}=\vec{s}_a+\vec{s}_b$. 
However, not all combinations of $\ell$ and $s$ are coupled to each other. For 
the conserving total angular momentum $J$ and the parity $\pi$, only few 
transitions of the type $(\ell,s)\leftrightarrow(\ell',s')$ are possible. 

In the present paper, we consider the low-energy ($E<3.4$\,MeV) collision of 
proton with the 
nucleus ${}^7\mathrm{Be}$. In its ground state, this nucleus has $J^\pi=3/2^-$. 
As a result of such a collision, the target nucleus may be excited to the state 
with $J^\pi=1/2^-$ at the energy $0.4291$\,MeV 
(see Ref.~\cite{Tilley_5_6_7}). The other excited states of 
${}^7\mathrm{Be}$ are too high as compared to the maximal collision energy 
and thus can be safely ignored. Therefore we deal with the following (elastic 
and inelastic) coupled processes:
\begin{equation}
\label{coupled_gammas}
\begin{array}{rlcl}
   \text{channel 1:\hspace{0.5cm}} &
   p+{}^7\mathrm{Be}(\frac32^-) &  & 
   p+{}^7\mathrm{Be}(\frac32^-)\\
   && \displaystyle\genfrac{}{}{0pt}{}{\searrow}{\nearrow}
   \bigcirc
     \displaystyle\genfrac{}{}{0pt}{}{\nearrow}{\searrow}
   & \\
   \text{channel 2:\hspace{0.5cm}} &
   p+{}^7\mathrm{Be}^*(\frac12^-) &  & 
   p+{}^7\mathrm{Be}^*(\frac12^-)\\
\end{array}\ ,
\end{equation}
where the circle in the middle is either the intermediate scattering state 
of the direct reaction or the compound resonance state of ${}^8\mathrm{B}$.
It is easy to see that the state $0^+$ of our eight-body system can only 
be formed if $\ell=1$ and $s=1$ in both channels. This means that we deal with 
a two-channel problem.

The  $N$-channel Jost matrices $\bm{f}^{\mathrm{(in)}}$ and 
$\bm{f}^{\mathrm{(out)}}$ are defined as the energy-dependent 
$(N\times N)$-``amplitudes'' of the incoming and 
outgoing multi-channel (diagonal-matrix) spherical waves, $\bm{H}^{(-)}$ and 
$\bm{H}^{(+)}$, in the asymptotic behaviour of the 
regular solution, $\bm{\phi}(E,r)$, of the radial Schr\"odinger equation,
\begin{equation}
\label{Jost_definition}
  \bm{\phi}(E,r)
  \mathop{\longrightarrow}\limits_{r\to\infty}
  \bm{H}^{(-)}(E,r)\bm{f}^{\mathrm{(in)}}(E)+
  \bm{H}^{(+)}(E,r)\bm{f}^{\mathrm{(out)}}(E)\ .
\end{equation}
A more detailed description of their meaning and properties can be found in 
Refs.~\cite{two_channel,our.MultiCh,our_Coulomb,our_He5}. It is worthwhile to 
write Eq.~(\ref{Jost_definition}) in the explicit form for the case of two 
coupled channels ($N=2$):
\begin{eqnarray}
\nonumber
  \bm{\phi}(E,r)
  &\mathop{\longrightarrow}\limits_{r\to\infty}&
  \begin{bmatrix}
  H_{\ell_1}^{(-)}(\eta_1,k_1r)e^{i\sigma_{\ell_1}} & 0\\[3mm]
  0 & H_{\ell_2}^{(-)}(\eta_2,k_2r)e^{i\sigma_{\ell_2}}
  \end{bmatrix}
  \begin{bmatrix}
  f_{11}^{\mathrm{(in)}}(E) & f_{12}^{\mathrm{(in)}}(E)\\[3mm]
  f_{21}^{\mathrm{(in)}}(E) & f_{22}^{\mathrm{(in)}}(E)
  \end{bmatrix}
  +\\[3mm]
\label{regular_ass}
  &+&
  \begin{bmatrix}
  H_{\ell_1}^{(+)}(\eta_1,k_1r)e^{-i\sigma_{\ell_1}} & 0\\[3mm]
  0 & H_{\ell_2}^{(+)}(\eta_2,k_2r)e^{-i\sigma_{\ell_2}}
  \end{bmatrix}
  \begin{bmatrix}
  f_{11}^{\mathrm{(out)}}(E) & f_{12}^{\mathrm{(in)}}(E)\\[3mm]
  f_{21}^{\mathrm{(out)}}(E) & f_{22}^{\mathrm{(in)}}(E)
  \end{bmatrix}\ ,
\end{eqnarray}
where
\begin{equation}
\label{Riccati_Coulomb}
   H_\ell^{(\pm)}(\eta,kr)=F_\ell(\eta,kr)\mp iG_\ell(\eta,kr)
   \ \mathop{\longrightarrow}\limits_{r\to\infty}
   \ \mp i\exp\left\{\pm i\left[kr-\eta\ln (2kr)
   -\frac{\ell\pi}{2}+\sigma_\ell\right]\right\}\ .
\end{equation}
In these equations, $k_n$, $\ell_n$, $\eta_n$, and $\sigma_{\ell_n}$ are the 
momentum, angular momentum, Sommerfeld parameter, and 
the pure Coulomb phase-shift in the channel $n$; the functions $F_\ell$ and 
$G_\ell$ are the standard regular and irregular Coulomb solutions of the 
Schr\"odinger equation (see, for example, Ref.~\cite{abramowitz}).

\subsection{Observables}
The $N$ columns of the matrix $\bm{\phi}(E,r)$ constitute a regular basis. 
Therefore a physical wave function, i.e. a column $\bm{u}(E,r)$, is their 
linear combination:
\begin{equation}
\label{physical_wf}
   \bm{u}(E,r) =\bm{\phi}(E,r)\bm{c}\ ,
\end{equation}
where $\bm{c}$ is a column matrix of the combination coefficients.
These coefficients are to be chosen to satisfy certain physical
boundary conditions at infinity. For a spectral point (either bound or a
resonant state) the physical wave function should only have the outgoing waves
in its asymptotic behaviour,
\begin{equation}
\label{spectral_bc}
  \bm{u}(E,r)
  \mathop{\longrightarrow}\limits_{r\to\infty}
  \bm{H}^{(-)}(E,r)
  \bm{f}^{\mathrm{(in)}}(E)\bm{c}
  +
  \bm{H}^{(+)}(E,r)
  \bm{f}^{\mathrm{(out)}}(E)\bm{c}\ .
\end{equation}
This can only be achieved if the first term in this equation is zero, i.e. if 
the unknown combination coefficients $c_n$ obey the homogeneous system of linear 
equations,
\begin{equation}
\label{fczero}
  \bm{f}^{\mathrm{(in)}}(E)\bm{c}=
  \begin{bmatrix}
  f_{11}^{\mathrm{(in)}}(E) & f_{12}^{\mathrm{(in)}}(E)\\[3mm]
  f_{21}^{\mathrm{(in)}}(E) & f_{22}^{\mathrm{(in)}}(E)
  \end{bmatrix}
  \begin{pmatrix} c_1 \\ c_2 \end{pmatrix}=0\ ,
\end{equation}
which has a non-zero solution if and only if
\begin{equation}
\label{detfzero}
  \det
  \begin{bmatrix}
  f_{11}^{\mathrm{(in)}}(E) & f_{12}^{\mathrm{(in)}}(E)\\[3mm]
  f_{21}^{\mathrm{(in)}}(E) & f_{22}^{\mathrm{(in)}}(E)
  \end{bmatrix}
  =0\ .
\end{equation}
The roots $E=\mathcal{E}_n$ of this equation are the spectral points. At real 
negative energies ($\mathcal{E}_n<0$) they
correspond to the bound states, and at the complex energies
($\mathcal{E}_n=E_r-i\Gamma/2$) they give us the resonances.

It is not difficult to shown (see, for example,
Refs.~\cite{our.MultiCh,two_channel}) that the scattering is determined by the
``ratio'' of the amplitudes of the out-going and in-coming waves, i.e. by the
$S$-matrix,
\begin{equation}
\label{Smatrix}
   \bm{S}(E)=\bm{f}^{\mathrm{(out)}}(E)
   \left[\bm{f}^{\mathrm{(in)}}(E)\right]^{-1}\ ,
\end{equation}
whose poles correspond to the roots of eq.~(\ref{detfzero}). 
The partial cross section that describes the transition between any two 
particular channels, can be obtained via the corresponding elements of 
the $S$-matrix (see, for example, Ref.~\cite{Frobrich}),
\begin{equation}
\label{particular_sigma}
   \sigma^J(n'\gets n)=\pi
   \frac{\mu_n k_{n'}}{\mu_{n'}k_n}\cdot
   \frac{2J+1}{2s+1}\left|
   \frac{S^J_{n'n}-\delta_{n'n}}{k_n}  
   \right|^2\ ,
\end{equation}
where $\mu_n$ is the reduced mass in the channel $n$.

The partial widths of a resonance can be found using the method
developed in Ref.~\cite{my.partial}:
\begin{equation}
\label{Gpartial}
   \Gamma_n=
   \displaystyle\frac{\mathrm{Re}(k_n)|\mathcal{A}_n|^2\Gamma}
   {\displaystyle\sum_{n'=1}^N\frac{\mu_n}{\mu_{n'}}
   \mathrm{Re}(k_{n'})|\mathcal{A}_{n'}|^2}\ ,
\end{equation}
where $\mathcal{A}_1$ and $\mathcal{A}_2$ are  the asymptotic amplitudes
(see Ref.~\cite{my.partial}) of the channels, given by
\begin{equation}
\label{A_1A_2_Final}
   \mathcal{A}_1=f^{(\mathrm{out})}_{11}-\displaystyle
   \frac{f^{(\mathrm{in})}_{11}f^{(\mathrm{out})}_{12}}{f^{(\mathrm{in})}_{12}}
   \ ,\qquad
   \mathcal{A}_2=f^{(\mathrm{out})}_{21}-\displaystyle
   \frac{f^{(\mathrm{in})}_{11}f^{(\mathrm{out})}_{22}}{f^{(\mathrm{in})}_{12}}
   \ .
\end{equation}
In these equations the Jost matrices are taken at the complex resonant energy.

\subsection{Analytic properties}
The Jost matrices (and thus the $S$-matrix) are multi-valued complex functions 
of the energy-variable $E$. They can be treated as single-valued, if considered 
on a multi-layered Riemann surface. Each thereshold is a branch point of such a 
surface. The multi-valuedness and thus the branching stem from the fact that 
the Jost matrices depend on the energy via the channel momenta,
\begin{equation}
\label{ch_momenta}
   k_n=\pm\sqrt{\frac{2\mu_n}{\hbar^2}(E-E_n)}\ ,
   \qquad  n=1,2,\dots,N\ ,
\end{equation}
where $E_n$ are the threshold energies. There are $2^N$ possible
combinations of the signs in front of the $N$ square roots (\ref{ch_momenta}), 
and thus for a single value of $E$ there are $2^N$ different values of the Jost 
matrices. If the interacting particles are charged, there is an additional 
uncertainty in calculating the Jost matrices for a given $E$. This is because 
the Coulomb spherical waves (\ref{Riccati_Coulomb}) and thus their amplitudes, 
$\bm{f}^{\mathrm{(in/out)}}(E)$, in the asymptotic behaviour 
(\ref{Jost_definition}) depend on the logarithms, $\ln k_n$, of the channel 
momenta. The complex function $\ln k_n$ has infinitely many different values,
\begin{eqnarray}
\label{Logarithm_ch_momenta}
   \ln k_n=\ln\left\{|k_n|e^{i[\arg(k_n)+2\pi m_n]}\right\} &=&
   \ln|k_n|+i\arg(k_n)+i2\pi m_n\ ,\\[3mm]
\nonumber
   && m_n = 0,\pm1,\pm2,\dots\ ,
\end{eqnarray}
corresponding to different choices of $m_n$. This implies that the Jost 
matrices are defined on a ``spiral'' Riemann surface with infinitely many layers 
(for more details see Ref.~\cite{our_He5}). At each threshold, this surface is
branching due to both the square-root and the logarithm multi-valuedness. The 
layers are identified by the signs of $\mathrm{Im}\,k_n$ and the logarithmic 
indices $m_n$. For the two-channel problem, the layers can be denoted by the 
symbols of the type $(\pm\pm)_{m_1m_2}$. The layers with $m_n\neq0$ are far 
away from the real axis, where the physical scattering energies belong to. This 
means that such layers may be safely ignored, and we should only consider the 
``principal'' layers corresponding to $m_1=m_2=0$. 

\begin{figure}
\centerline{\epsfig{file=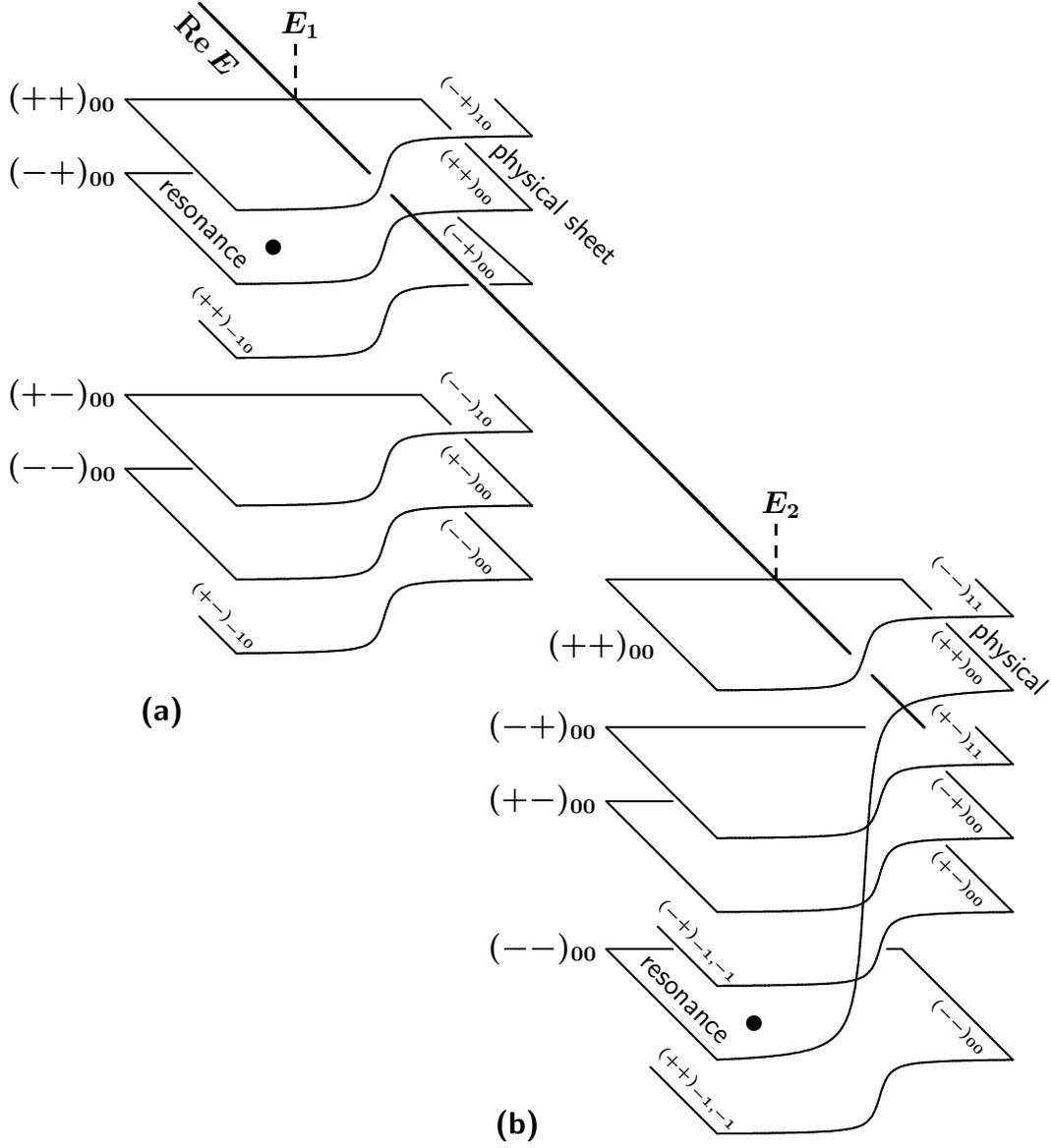}}
\caption{\sf
Interconnections of the Riemann sheets for the two-channel problem, where the
Coulomb potential is present in both channels:
(a) the interconnections between the thresholds $E_1$ and $E_2$; (b) the
interconnections above the highest threshold. The symbols
$(\pm\pm)_{m_1m_2}$ label the sheets where $\mathrm{Im}(k_1)$ and 
$\mathrm{Im}(k_2)$
are either positive or
negative, and the subscripts $m_1m_2$  are the numbers of $i2\pi$ in
Eq.~(\protect{\ref{Logarithm_ch_momenta}}) for the channels.
}
\label{fig.sheets_2ch_Coulomb}
\end{figure}

\begin{figure}
\centerline{\epsfig{file=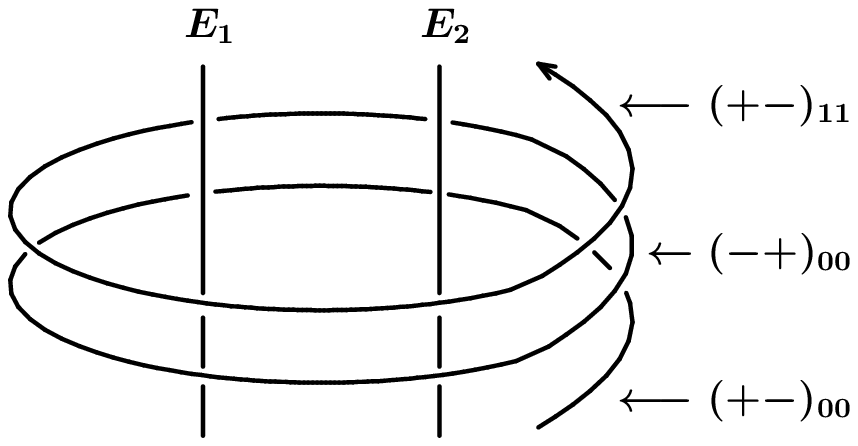}}
\caption{\sf
Two-circle path around both thresholds ($E_1$ and $E_2$) on the Riemann surface 
of a two-channel problem with the Coulomb forces in both channels. If the 
starting point is on the sheet $(+-)_{00}$, then the final point is on the 
sheet $(+-)_{11}$. 
}
\label{fig.spiral.elliptic}
\end{figure}

For our two-body problem, the Riemann surface is schematically depicted in 
Fig.~\ref{fig.sheets_2ch_Coulomb}. Each sheet of this surface is cut along 
its own real axis and the interconnections among the cuts are done in such a 
way that one full circle around the threshold $E_n$ changes the sign of 
$\mathrm{Im}\,k_n$, two full circles around $E_n$ change the logarithmic index 
$m_n$. If we go around both thresholds, then both momenta and both logarithmic 
indices do change. This is illustrated in Fig.~\ref{fig.spiral.elliptic}.

\subsection{Analytic structure}
The multi-channel Riemann surface has a very complicated topology. 
The intricate interconnections of its layers should be kept in mind not only in 
pure theoretical considerations, but also in the analysis of experimental data 
when one tries to extract the information on the resonances. The reason is that 
the resonance poles of the $S$-matrix lie on one of the Riemann sheets. In 
order to reach them, one has to do the analytic continuation of the $S$-matrix, 
starting from the real axis. In doing this, one should be careful, and 
especially when such a continuation is done near a branch point.

All the complications caused by the branching of the Riemann surface, can be 
circumvented by using the semi-analytic representations of the Jost matrices 
suggested in Refs.~\cite{our.MultiCh, our_Coulomb}. In these representations, 
the factors responsible for the branching of the Riemann surface are given 
explicitly. For the charged-particle case, it was shown\cite{our_Coulomb} that 
the Jost matrices have the following structure
\begin{equation}
\label{multi.Finout_structure}
   \bm{f}^{(\mathrm{in/out})}=
   \bm{Q}^{(\pm)}\left[
   \bm{D}^{-1}\bm{A}\bm{D}-(\bm{M}\pm i)
   \bm{K}^{-1}\bm{D}\bm{B}\bm{D}\right]\ ,   
\end{equation}
where the unknown matrices $\bm{A}(E)$ and $\bm{B}(E)$ are single-valued 
functions of the energy and are defined on the simple energy-plane without any 
branch points. All the troubles with the branching stem from the explicitly 
given factors (diagonal matrices):
\begin{equation}
\label{multi.defQ}
   \bm{Q}^{(\pm)} =
   \operatorname{diag}\left\{
   \frac{e^{\pi\eta_1/2}\ell_1!}{\Gamma(\ell_1+1\pm i\eta_1)},
   \frac{e^{\pi\eta_2/2}\ell_2!}{\Gamma(\ell_2+1\pm i\eta_2)},\dots,
   \frac{e^{\pi\eta_N/2}\ell_N!}{\Gamma(\ell_N+1\pm i\eta_N)}\right\}\ ,
\end{equation}

\begin{equation}
\label{multi.defD}
   \bm{D} =
   \operatorname{diag}\left\{
   C_{\ell_1}(\eta_1)k_1^{\ell_1+1},
   C_{\ell_2}(\eta_2)k_2^{\ell_2+1},\dots,
   C_{\ell_N}(\eta_N)k_N^{\ell_N+1}\right\}\ ,
\end{equation}

\begin{equation}
   \bm{M} =
   \operatorname{diag}\left\{
   \frac{2\eta_1h(\eta_1)}{C_0^2(\eta_1)},
   \frac{2\eta_2h(\eta_2)}{C_0^2(\eta_2)},
   \dots,
   \frac{2\eta_Nh(\eta_N)}{C_0^2(\eta_N)}\right\}\ ,
\end{equation}

\begin{equation}
\label{multi.chmom}
   \bm{K}=\operatorname{diag}\left\{k_1,k_2,\dots,k_N\right\}\ .
\end{equation}
They involve the Coulomb barrier
factor $C_\ell$ and the function $h(\eta)$ that is responsible for the 
logarithmic branching:
\begin{equation}
\label{CL}
  C_\ell(\eta)=
  \frac{2^\ell e^{-\pi\eta/2}}{(2\ell)!!}
  \exp\left\{\frac12\left[\ln\Gamma(\ell+1+i\eta)+
  \ln\Gamma(\ell+1-i\eta)\right]\right\}
  \ \mathop{\longrightarrow}\limits_{\eta\to0}\ 1\ ,
\end{equation}
\begin{equation}
\label{h_function}
   h(\eta)=\frac12\left[\psi(i\eta)+
   \psi(-i\eta)\right]-\ln{\eta}\ ,
   \qquad
   \psi(z)=\frac{\Gamma'(z)}{\Gamma(z)}\ ,
   \qquad
   {\eta}=\frac{e^2Z_1Z_2\mu}{\hbar^2k}\ .
\end{equation}
In the explicit form for the matrix elements, 
Eq.~(\ref{multi.Finout_structure}) can be written as
\begin{eqnarray}
\label{multi.matrixelements}
   f^{(\mathrm{in/out})}_{mn}(E) &=&
   \frac{e^{\pi\eta_m/2}\ell_m!}{\Gamma(\ell_m+1\pm i\eta_m)}
   \left\{
   \frac{{C}_{\ell_n}(\eta_n)k_n^{\ell_n+1}}
   {{C}_{\ell_m}(\eta_m)k_m^{\ell_m+1}}{A}_{mn}(E)\ -\right.\\[3mm]
\nonumber
   &-&
   \left.\left[
   \frac{2\eta_mh(\eta_m)}{C_0^2(\eta_m)}\pm i\right]
   {C}_{\ell_m}(\eta_m){C}_{\ell_n}(\eta_n)
   k_m^{\ell_m}k_n^{\ell_n+1}{B}_{mn}(E)\right\}\ .
\end{eqnarray}
The matrices $\bm{A}(E)$ and $\bm{B}(E)$ are
the same for both $\bm{f}^{\rm(in)}$ and $\bm{f}^{\rm(out)}$, and they are real
for real energies.

Apparently, the analytic structure of the $S$-matrix (\ref{Smatrix}) is even 
more complicated than that of the Jost matrices. This means that none of the 
simplified phenomenological formulae for the multi-channel $S$-matrix (that 
very often are used to fit experimental data) can guarantee the correct 
topology of the Riemann surface. The consequences of such a simplification for 
the analytic continuation of the $S$-matrix are unclear and unpredictable.

\subsection{Approximation and analytic continuation}
\label{sect.ApprCont}
In the exact expressions (\ref{multi.matrixelements}), the only unknowns are the
matrices $\bm{A}(E)$ and $\bm{B}(E)$, which are single-valued and analytic. 
They can be expanded in Taylor series around an arbitrary complex
energy $E_0$,
\begin{equation}
\label{A.Taylor}
   \bm{A}(E)=\bm{a}^{(0)}+\bm{a}^{(1)}(E-E_0)+
   \bm{a}^{(2)}(E-E_0)^2+\cdots\ ,
\end{equation}
\begin{equation}
\label{B.Taylor}
   \bm{B}(E)=\bm{b}^{(0)}+\bm{b}^{(1)}(E-E_0)+
   \bm{b}^{(2)}(E-E_0)^2+\cdots\ .
\end{equation}
Here $\bm{a}^{(m)}(E_0)$ and $\bm{b}^{(m)}(E_0)$ are the $(N\times
N)$-matrices (for a two-channel case, $N=2$) depending on the choice of the 
center $E_0$ of the expansion. The matrix elements of $\bm{a}^{(m)}(E_0)$ and 
$\bm{b}^{(m)}(E_0)$ are the
unknown parameters. We can take the first several terms of these
expansions and find the unknown parameters by
fitting some available experimental data. As a result, we obtain 
approximate analytic expressions (\ref{multi.matrixelements}) for the Jost 
matrices.

\begin{figure}
\centerline{\epsfig{file=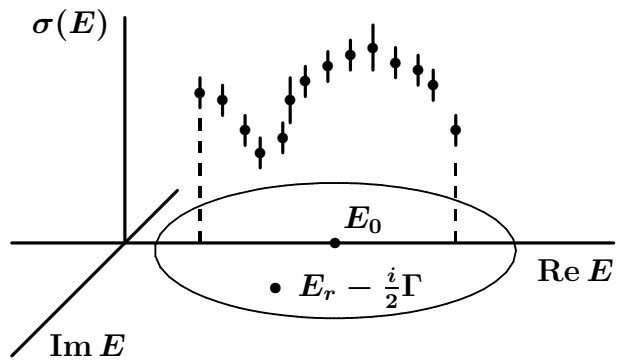}}
\caption{\sf
The data points fitted within an interval centered at $E_0$ on the real axis, 
give us the Jost matrices valid within a circle of the adjacent complex domain, 
where the resonance spectral points can be found.
}
\label{fig.fit.around.E0}
\end{figure}

It is convenient to choose the central point $E_0$ on the real axis. Such a 
choice makes the matrices $\bm{a}^{(m)}(E_0)$ and $\bm{b}^{(m)}(E_0)$ real. 
After adjusting the parameters (via fitting the data) we can consider the same 
Jost matrices (\ref{multi.matrixelements}) at complex energies and thus can 
locate the resonances, as is schematically illustrated in 
Fig.~\ref{fig.fit.around.E0}.

When looking at complex $E$, we can choose the
appropriate sheet of the Riemann surface. The single-valued functions
$\bm{A}(E)$ and $\bm{B}(E)$ are the same on all the sheets. The differencies
only stem from the explicit factors depending on $k_n$ and $\ln k_n$ in
Eq.~(\ref{multi.matrixelements}).
For a given energy $E$, we calculate the square roots (\ref{ch_momenta}) and 
$\ln k_n$ for all the channel
momenta. Choosing appropriate signs in front of the square roots and adding 
appropriate number of $2\pi$ in Eq.~(\ref{Logarithm_ch_momenta}), we can place 
the point on any Riemann sheet that we need. In other words, the analytic 
continuation of the Jost matrices from the real axis (where the fitting is done) 
to a chosen Riemann sheet is always done correctly despite the approximations
(\ref{A.Taylor},\ref{B.Taylor}).

\section{The data and fitting procedure}
\label{sec.fitting}
The Jost matrices describe the two-body states with definite quantum numbers, 
namely, $(J^\pi,\ell,s)$. If we were trying to fit ``raw'' experimental data, 
we would need to sum up several states with different $J$ and many partial 
waves. This would result in too many free parameters and the task would 
become unmanageable. To avoid such a difficulty, we consider partial cross 
sections (for a given $J^\pi$) separately. 

In the present work, we deal with the  state $0^+$  of the system
$p\,{}^7\mathrm{Be}$. This state involves only one partial wave, 
namely, $(\ell,s)=(1,1)$ in both channels. In order to obtain the partial cross 
sections, one has to do the partial-wave analysis of the ``raw'' data. 
This is a very complicated task by itself. We therefore rely on the existing 
$R$-matrix analysis of the system
$p\,{}^7\mathrm{Be}$, published in Ref.~\cite{Rogachev2013}, where the 
experimental data on the 
excitation functions for the elastic and inelastic $p\,{}^7\mathrm{Be}$ 
scattering were parametrized. As a result of this analysis, the authors of 
Ref.~\cite{Rogachev2013} reported three new low-energy resonances with the 
quantum numbers $0^+$, $1^+$ and $2^+$.

Using the parameters given in Ref.~\cite{Rogachev2013}, we construct the 
$R$-matrix and then the corresponding $S$-matrix, 
from which any partial cross section can be calculated. Since the $R$-matrix 
of Ref.~\cite{Rogachev2013} was obtained by fitting the ``raw'' data, the 
partial cross sections we obtain from this $R$-matrix, can be considered as 
experimental. In a sense, such an approach is similar to treating the 
scattering phase-shifts as experimental data despite the fact that nobody 
measures them directly and they are obtained from a complicated partial-wave
analysis of the ``raw'' data.

\begin{figure}
\centerline{\epsfig{file=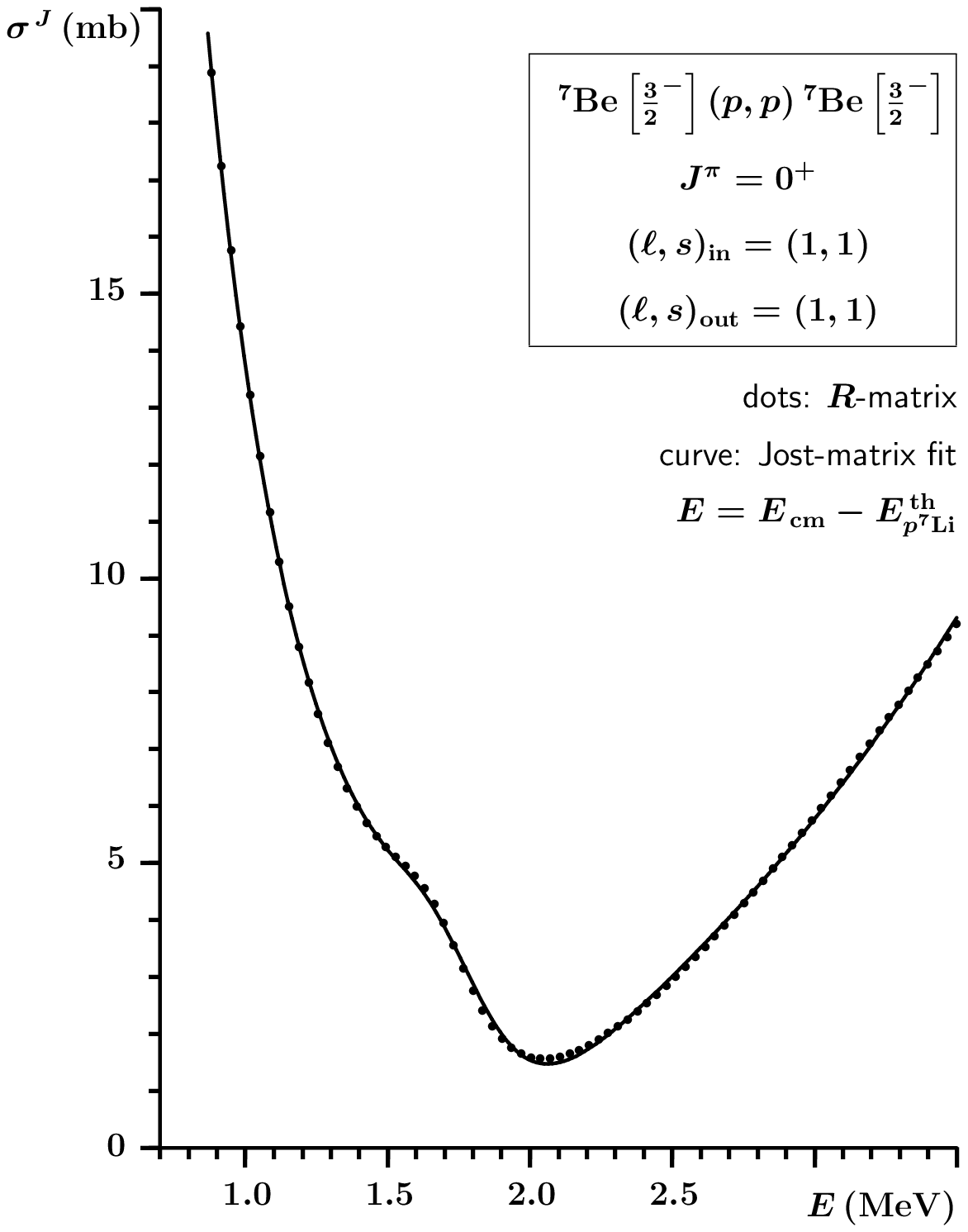}}
\caption{\sf
Partial cross section for the transition $1\to1$ of the processes 
(\ref{coupled_gammas}) in the state with $J^\pi=0^+$. The dots are the 
``experimental'' points obtained from the $R$-matrix taken from 
Ref.~\cite{Rogachev2013}. The curve is our fit with the Jost matrix parameters 
given in Table~\ref{table.parameters}. The collision energy is counted from the 
$p\,{}^7\mathrm{Be}$ threshold.
}
\label{fig.fit.11}
\end{figure}

\begin{figure}
\centerline{\epsfig{file=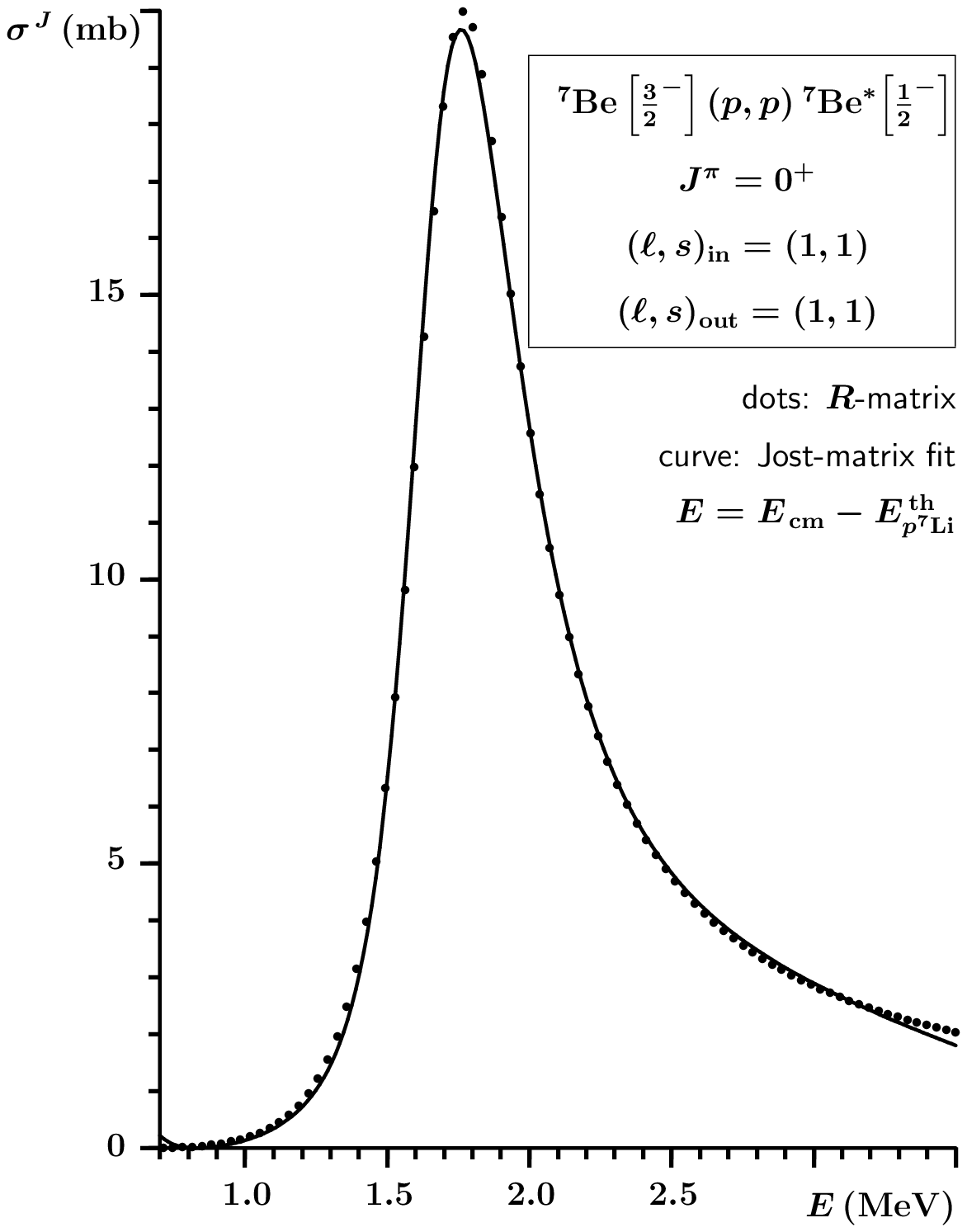}}
\caption{\sf
Partial cross section for the transition $1\to2$ of the processes 
(\ref{coupled_gammas}) in the state with $J^\pi=0^+$. The dots are the 
``experimental'' points obtained from the $R$-matrix taken from 
Ref.~\cite{Rogachev2013}. The curve is our fit with the Jost matrix parameters 
given in Table~\ref{table.parameters}. The collision energy is counted from the 
$p\,{}^7\mathrm{Be}$ threshold.
}
\label{fig.fit.12}
\end{figure}

\begin{figure}
\centerline{\epsfig{file=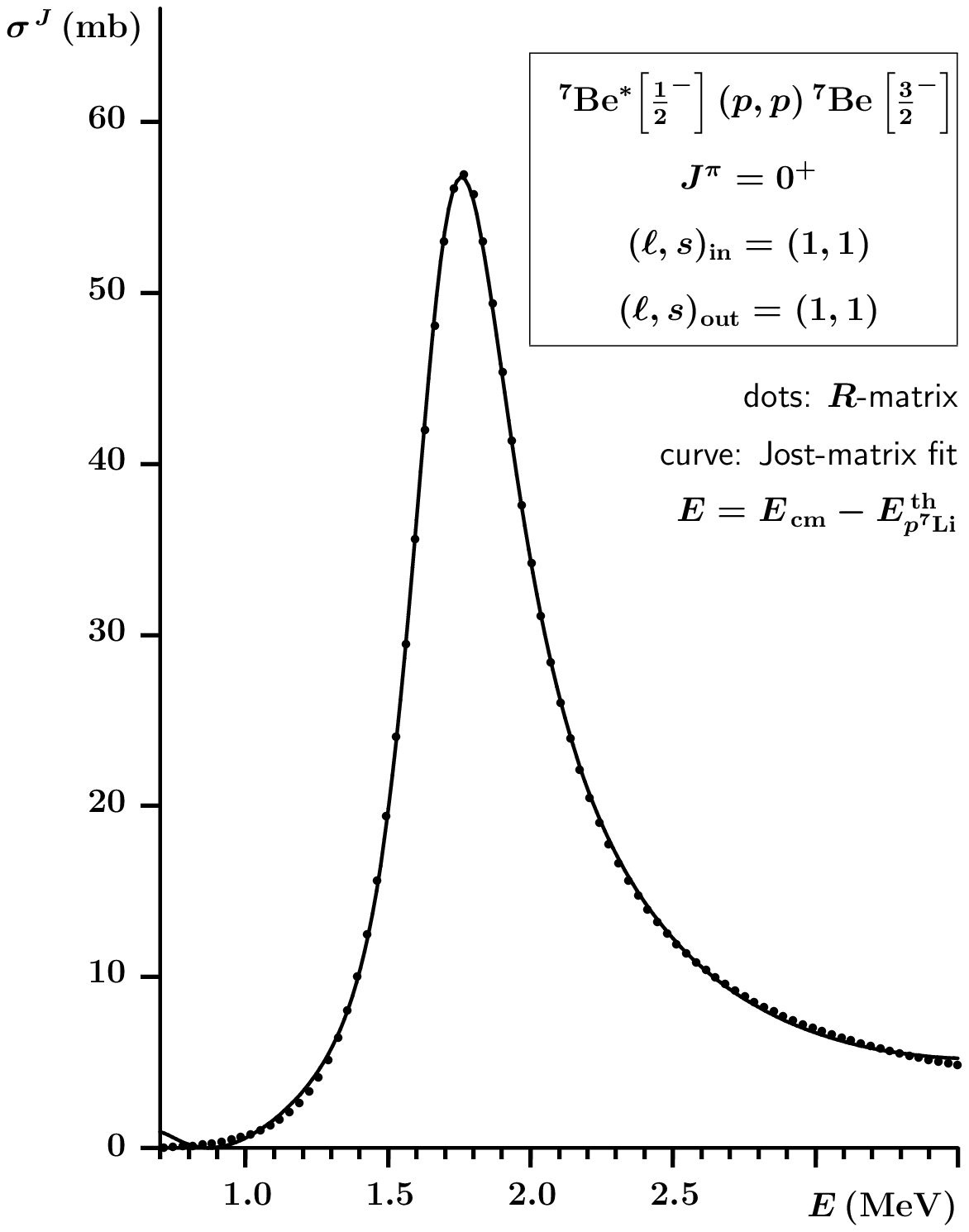}}
\caption{\sf
Partial cross section for the transition $2\to1$ of the processes 
(\ref{coupled_gammas}) in the state with $J^\pi=0^+$. The dots are the 
``experimental'' points obtained from the $R$-matrix taken from 
Ref.~\cite{Rogachev2013}. The curve is our fit with the Jost matrix parameters 
given in Table~\ref{table.parameters}. The collision energy is counted from the 
$p\,{}^7\mathrm{Be}$ threshold.
}
\label{fig.fit.21}
\end{figure}

\begin{figure}
\centerline{\epsfig{file=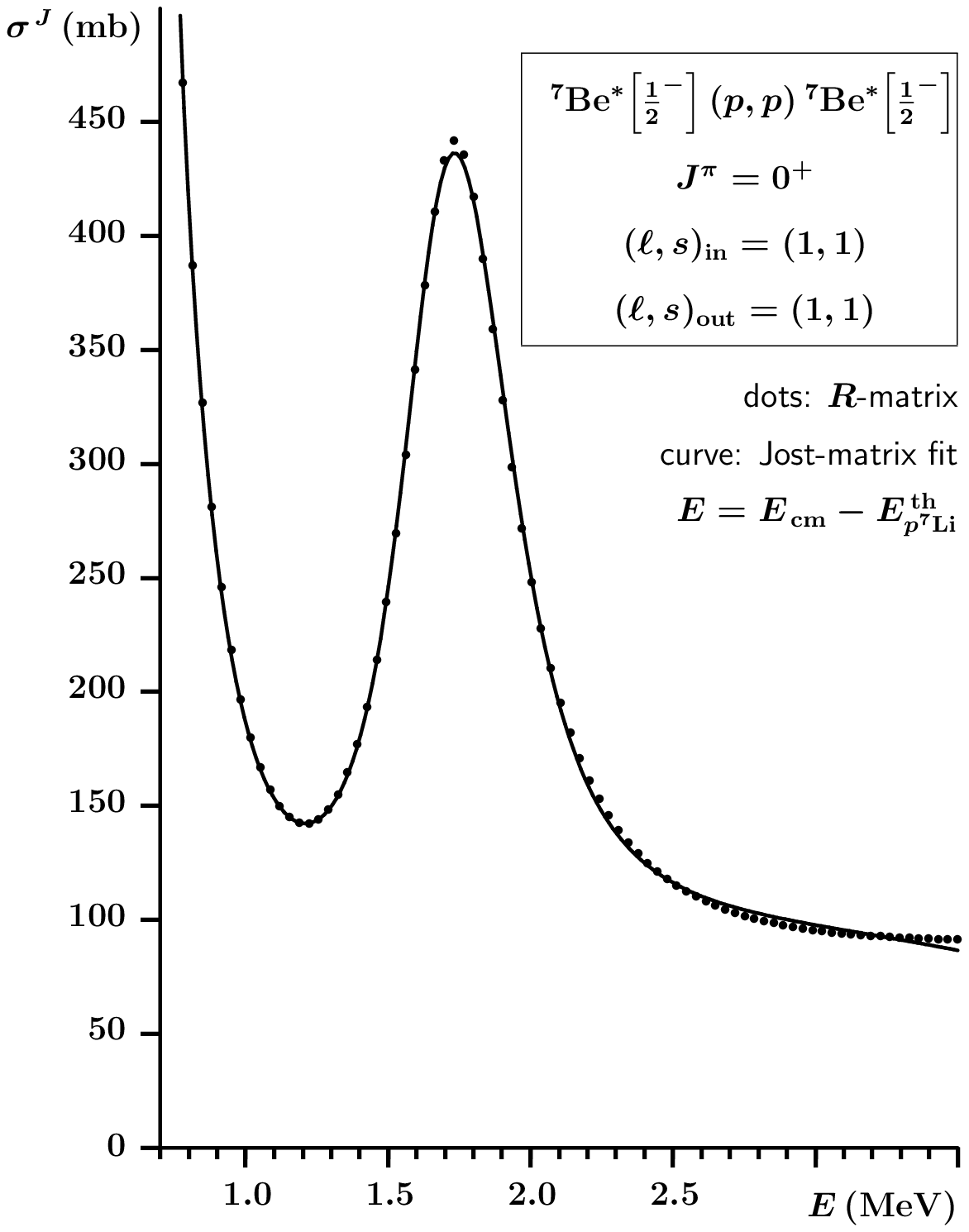}}
\caption{\sf
Partial cross section for the transition $2\to2$ of the processes 
(\ref{coupled_gammas}) in the state with $J^\pi=0^+$. The dots are the 
``experimental'' points obtained from the $R$-matrix taken from 
Ref.~\cite{Rogachev2013}. The curve is our fit with the Jost matrix parameters 
given in Table~\ref{table.parameters}. The collision energy is counted from the 
$p\,{}^7\mathrm{Be}$ threshold.
}
\label{fig.fit.22}
\end{figure}

Thus obtained partial cross sections for the four processes 
(\ref{coupled_gammas}) are given in Figs.~\ref{fig.fit.11}-\ref{fig.fit.22}, 
where they are shown by the dots. We 
consider these dots as the (indirectly obtained) experimental points, which we 
fit by varying the Jost matrices.

As the basis for parametrizing the Jost matrices, we use the semi-analytic 
expressions (\ref{multi.matrixelements}), where the unknown matrices 
$\bm{A}(E)$ and $\bm{B}(E)$ are analytic and single-valued functions of 
the energy. We therefore can approximate them by the $(M+1)$ Taylor terms,
\begin{eqnarray}
\label{Aapprox}
   A_{n'n}(E) &\approx& \sum_{m=0}^M a^{(m)}_{n'n}(E-E_0)^m\ ,\\[3mm]
\label{Bapprox}
   B_{n'n}(E) &\approx& \sum_{m=0}^M b^{(m)}_{n'n}(E-E_0)^m\ ,
   \qquad n',n=1,2,\dots,N\ ,
\end{eqnarray}
with $E_0$ taken somewhere in the middle of the interval covered by the 
experimental points. The unknown expansion coefficients $a^{(m)}_{n'n}$ and 
$b^{(m)}_{n'n}$ serve as the fitting parameters and $N=2$
is the number of the coupled channels. These matrices $\bm{A}$ and $\bm{B}$, 
when substituted in the semi-analytic expressions
(\ref{multi.matrixelements}), give us the approximate Jost matrices and the 
corresponding $S$-matrix (\ref{Smatrix}), which is used to calculate the 
approximate partial cross sections (\ref{particular_sigma}), 
$\tilde{\sigma}_{n'\gets n}$, depending on the fitting parameters.

\begin{table}
\begin{center}
\begin{tabular}{|c|c|c|c|c|c|}
\hline
\multicolumn{3}{|c|}{$E_0$}
& $1.6\,\mathrm{MeV}$ & $1.8\,\mathrm{MeV}$ & $2.0\,\mathrm{MeV}$\\
\hline
$m$ & $n'$ & $n$
&
\parbox{2cm}{\begin{center}
$a_{n'n}^{(m)}$,\ $b_{n'n}^{(m)}$\\[1mm]
$[\mathrm{MeV}^{-m}]$\end{center}}
&
\parbox{2cm}{\begin{center}
$a_{n'n}^{(m)}$,\ $b_{n'n}^{(m)}$\\[1mm]
$[\mathrm{MeV}^{-m}]$\end{center}}
&
\parbox{2cm}{\begin{center}
$a_{n'n}^{(m)}$,\ $b_{n'n}^{(m)}$\\[1mm]
$[\mathrm{MeV}^{-m}]$\end{center}}\\
\hline
\multirow{4}{*}{0} & 1 & 1 &
$        -3.8980$,\   $          260.02    $ &
$        -2.5158$,\   $          132.16    $ &
$        -1.6501$,\   $         51.674    $ \\
  & 1 & 2 &
$        -0.076997$,\   $         19.287    $ &
$        -0.85222$,\   $         61.581    $ &
$        -1.3865$,\   $         66.164    $ \\
  & 2 & 1 &
$         0.0071846$,\   $          142.56    $ &
$         0.28927$,\   $          113.93    $ &
$         0.46568$,\   $         71.761    $ \\
  & 2 & 2 & 
$        -0.19860$,\   $        -19.181    $ &
$        -0.16232$,\   $         13.827    $ &
$        -0.10506$,\   $         29.145    $ \\
\hline
\multirow{4}{*}{1} & 1 & 1 &
$         3.5822$,\   $         -537.13    $ &
$        -2.4338$,\   $        -67.781    $ &
$        -5.7215$,\   $         37.455    $ \\
 & 1 & 2 &
$         0.18923$,\   $        -15.205    $ &
$        -0.74242$,\   $        -1.6230$ &
$        -4.5979$,\   $         69.555    $ \\
 & 2 & 1 & 
$         1.2508$,\   $         41.701    $ &
$         2.0021$,\   $          149.42    $ &
$         2.0154$,\   $          131.71    $\\
 & 2 & 2 &
$        -0.036673$,\   $        -13.918    $ &
$        -0.028947$,\   $         22.033    $ &
$        -0.37030$,\   $         42.200    $ \\
\hline
\multirow{4}{*}{2} & 1 & 1 &
$        -7.7575$,\   $          1313.9    $ &
$        -19.442    $,\   $          581.08    $ &
$        -16.051    $,\   $          118.57    $ \\
 & 1 & 2 & 
$        -0.46898$,\   $          111.02    $ &
$        -6.3748$,\   $          250.95    $ &
$        -12.680    $,\   $          135.59    $ \\
 & 2 & 1 &
$         2.8715$,\   $          353.10    $ &
$         3.7550$,\   $          330.22    $ &
$         3.4106$,\   $          192.31    $ \\
 & 2 & 2 & 
$        -0.77670$,\   $         10.873    $ &
$        -1.3155$,\   $         33.713    $ &
$        -1.8707$,\   $         25.314    $ \\ 
\hline
\multirow{4}{*}{3} & 1 & 1 &
$        -29.889    $,\   $         -698.57    $ &
$        -27.627    $,\   $         -462.76    $ &
$        -14.980    $,\   $         -195.76    $ \\
 & 1 & 2 & 
$         0.032721$,\   $        -78.511    $ &
$        -8.0587$,\   $         -198.45    $ &
$        -10.889    $,\   $         -190.90    $ \\
 & 2 & 1 & 
$         2.8100$,\   $          220.84    $ &
$         3.0660$,\   $          109.69    $ &
$         2.6138$,\   $         57.036    $ \\
 & 2 & 2 & 
$        -2.4411$,\   $         -151.14    $ &
$        -2.5425$,\   $        -92.968    $ &
$        -2.2098$,\   $        -51.518    $ \\
\hline
\multirow{4}{*}{4} & 1 & 1 &
$        -13.780    $,\   $        -18.493    $ &
$        -9.4436$,\   $        -19.098    $ &
$        -4.1096$,\   $        -12.805    $ \\
 & 1 & 2 & 
$         0.86303$,\   $         27.495    $ &
$        -2.1777$,\   $         19.390    $ &
$        -2.5535$,\   $         14.865    $ \\
 & 2 & 1 & 
$         1.2247$,\   $        -79.035    $ &
$         0.98110$,\   $        -50.869    $ &
$         0.74022$,\   $        -30.866    $ \\
 & 2 & 2 & 
$        -1.4628$,\   $         26.744    $ &
$        -1.1201$,\   $         10.597    $ &
$        -0.72228$,\   $        -1.1996$ \\
\hline
\end{tabular}
\end{center}
\caption{\sf
Parameters of the expansions (\ref{Aapprox},\ref{Bapprox}) with three choices 
of the central point $E_0$. These
parameters  for $E_0=1.8$\,MeV were used to generate the curves shown in Figs.
\ref{fig.fit.11}, \ref{fig.fit.12}, \ref{fig.fit.21}, and \ref{fig.fit.22}.
}
\label{table.parameters}
\end{table}

The optimal values of the fitting parameters are found by minimizing the
following function:
\begin{eqnarray}
\label{chisquare}
   \chi^2 &=&  
    W_{11} \sum_{i=1}^K\left|
    \tilde{\sigma}_{1\gets 1}(E_i)-\sigma_{1\gets 1}(E_i)\right|^2+\\[3mm]
\nonumber
    &+&
    W_{21} \sum_{i=1}^K\left|
    \tilde{\sigma}_{2\gets 1}(E_i)-\sigma_{2\gets 1}(E_i)\right|^2+\\[3mm]
\nonumber 
    &+&
    W_{12} \sum_{i=1}^K\left|
    \tilde{\sigma}_{1\gets 2}(E_i)-\sigma_{1\gets 2}(E_i)\right|^2+\\[3mm]
\nonumber 
    &+&
    W_{22} \sum_{i=1}^K\left|
    \tilde{\sigma}_{2\gets 2}(E_i)-\sigma_{2\gets 2}(E_i)\right|^2\ , 
\end{eqnarray}
where $K$ is the number of experimental points, and $\sigma_{n'\gets n}(E_i)$ is 
the experimental cross section at the energy $E_i$. The experimental
errors are not defined because the data are taken from the $R$-matrix analysis. 
We therefore put all of them to unity in the $\chi^2$-function 
(\ref{chisquare}). Since the experimental errors are absent, each point is 
equally important in this function. However the magnitudes of the cross 
sections in different channels are significantly different (compare, for 
example, Figs. \ref{fig.fit.11} and   \ref{fig.fit.22}). As a result of such a 
difference, the minimization tends to give preference to the curves with larger 
values of $\sigma_{n'\gets n}$, while the quality of the fitting of the smaller 
cross sections remains poor. To avoid this tendency, we introduce the weight 
factors $W_{n'n}$ in the $\chi^2$-function (\ref{chisquare}). These factors are 
chosen in such a way that the contributions from the four terms are more or 
less the same.

For the minimization, we use the MINUIT program developed in
CERN~\cite{MINUIT}. The function (\ref{chisquare}) has many local minima. 
The search for the best of them can be based on the following strategy. First 
of all, the minimization procedure should be repeated many times (we did it 
$\sim1000$ times) with randomly chosen starting values of the parameters. Then, 
after a good minimum is found, it can be refined by choosing random starting 
point around the best point found in the parameter space. After each 
improvement, the new starting parameters are chosen by random variations 
of the parameters around the new best point.

The cross sections (as well as any other observables) are expressed via the 
elements of the $S$-matrix (\ref{Smatrix}), i.e. via the ratio of the Jost 
matrices. In such a ratio, any common factor in $\bm{f}^{(\mathrm{in})}$ 
and $\bm{f}^{(\mathrm{out})}$ cancels out. This means that the set 
of the parameters $a^{(m)}_{n'n}$ and $b^{(m)}_{n'n}$ can be scaled by any 
convenient factor. Such a scaling does not affect any results.

\section{Results}
\label{sec.results}
The experimental data (obtained from the $R$-matrix given in
Ref.~\cite{Rogachev2013}) for the four processes (\ref{coupled_gammas}), were 
fitted as is decribed in Sec.~\ref{sec.fitting}, with $M=4$,
$W_{11}=3$, $W_{12}=15$, $W_{21}=0.04$, and $W_{22}=0.002$. We repeated the fit 
with five different values of the central energy $E_0$, namely, with
$E_0=1.6$\,MeV, $E_0=1.7$\,MeV, $E_0=1.8$\,MeV, $E_0=1.9$\,MeV, 
and $E_0=2.0$\,MeV (the energy is counted from the 
$p\,{}^7\mathrm{Be}$-threshold).  Formally, the results should not depend on 
the choice of $E_0$. However, the Taylor expansions 
(\ref{Aapprox}, \ref{Bapprox}) are truncated and the minimization procedure 
is always an approximate one. The calculations with several different $E_0$ 
allow us to see how stable the results are, to find the average values of the 
resonance parameters and their standard deviations, and to exclude any possible 
spurious poles of the $S$-matrix (that should be unstable).

The results of the fit with $E_0=1.8$\,MeV are
graphically shown in Figs. \ref{fig.fit.11}, \ref{fig.fit.12}, \ref{fig.fit.21},
and \ref{fig.fit.22}. For the other choices of $E_0$, the quality of the fit is 
the same and it would be impossible to distinguish the corresponding curves. 
The optimal parameters for the three (out of five) choices of $E_0$ are given 
in Table~\ref{table.parameters}. The units for the parameters are chosen in 
such a way that the Jost matrices are dimensionless.

The resonances were sought as zeros of $\det\bm{f}^{\mathrm{(in)}}(E)$ on the 
principal sheet $(--)_{00}$ of the Riemann surface. This was done using the 
Newton's method~\cite{Press}. In this way, we found {\it two resonances that are 
close to each other}. For each of the five choices of $E_0$, their parameters 
are given in Tables~\ref{table.resonance1} and \ref{table.resonance2}. It is 
seen that our procedure gives at least three stable digits.

\begin{table}
\begin{center}
\begin{tabular}{|c|c|c|c|c|}
\hline
$E_0$\,(MeV) & $E_r$\,(MeV) & $\Gamma$\,(MeV) & $\Gamma_{1}$\,(MeV) &
$\Gamma_{2}$\,(MeV)\\
\hline
$1.6$ &
$ 1.65255$ &
$ 0.44772$ &   
$ 0.13420$ &    
$ 0.31352$ \\
\hline
$1.7$ &
$1.65198$ &   
$0.44653$ &        
$0.13357$ &         
$0.31295$ \\
\hline
$1.8$ &
$1.65283$ &    
$0.44908$ &    
$0.13486$ &   
$0.31422$ \\
\hline
$1.9$ &
$1.65230$ &
$0.44713$ &    
$0.13404$ &     
$0.31309$ \\
\hline 
$2.0$ &
$1.65223$ &
$0.44745$ &    
$0.13412$ &    
$0.31333$ \\
\hline
\end{tabular}
\end{center}
\caption{\sf
Parameters of the first $0^+$ resonance found with five different choices of 
the expansion parameter $E_0$. The energy $E_r$ is counted from the 
$p\,{}^7\mathrm{Be}$ threshold. The partial widths $\Gamma_1$ and $\Gamma_2$ 
correspond to the elastic and inelastic channels, respectively.
}
\label{table.resonance1}
\end{table}

\begin{table}
\begin{center}
\begin{tabular}{|c|c|c|c|c|}
\hline
$E_0$\,(MeV) & $E_r$\,(MeV) & $\Gamma$\,(MeV) & $\Gamma_{1}$\,(MeV) &
$\Gamma_{2}$\,(MeV)\\
\hline
$1.6$ &
$1.81879$ &
$0.83658$ &     
$0.54461$ &    
$0.29196$ \\
\hline
$1.7$ &
$1.82066$ &
$0.83932$ &       
$0.53713$ &       
$0.30219$ \\
\hline
$1.8$ &
$1.81891$ &   
$0.84133$ &    
$0.54164$ &    
$0.29968$ \\
\hline
$1.9$ & 
$1.81947$ &
$0.86394$ &    
$0.51899$ &     
$0.34495$ \\
\hline 
$2.0$ &
$1.82101$ &
$0.83255$ &   
$0.55292$ &    
$0.27963$ \\
\hline
\end{tabular}
\end{center}
\caption{\sf
Parameters of the second $0^+$ resonance found with five different choices of 
the expansion parameter $E_0$. The energy $E_r$ is counted from the 
$p\,{}^7\mathrm{Be}$ threshold. The partial widths $\Gamma_1$ and $\Gamma_2$ 
correspond to the elastic and inelastic channels, respectively.
}
\label{table.resonance2}
\end{table}

\begin{table}
\begin{center}
\begin{tabular}{|c|c|c|c|c|}
\hline
$E_{\mathrm{ex}}$\,(MeV) & $\Gamma$\,(MeV) & $\Gamma_{1}$\,(MeV) & 
$\Gamma_{2}$\,(MeV) & \sffamily{Ref.}\\
\hline
$1.7899\pm0.0003$ & 
$0.4476\pm0.0009$ &
$0.1342\pm0.0005$ &
$0.3134\pm0.0005$ &
\sffamily{this work}\\
\hline
$1.9573\pm0.0010$ &
$0.8427\pm0.0123$ &
$0.5391\pm0.0126$ &
$0.3037\pm0.0247$ &
\sffamily{this work}\\
\hline
$1.9\pm0.1$ &
$0.53\begin{array}{cl}+&0.60\\[-2mm] -&0.10\end{array}$ &
$0.06\begin{array}{cl}+&0.30\\[-2mm] -&0.02\end{array}$ &
$0.47\begin{array}{cl}+&0.40\\[-2mm] -&0.10\end{array}$ &
\cite{Rogachev2013}\\
\hline
\end{tabular}
\end{center}
\caption{\sf
Statistically averaged parameters of the two $0^+$ resonances (the first two 
lines) and the single $0^+$ resonance reported in Ref.~\cite{Rogachev2013}. The 
energy $E_{\mathrm{ex}}$ is counted from the ground state of ${}^8\mathrm{B}$ 
nucleus. 
}
\label{table.resonances.average}
\end{table}

\begin{table}
\begin{center}
\begin{tabular}{|c|c|c|c|}
\hline
\multicolumn{4}{|c|}{\sffamily $S$-matrix poles (MeV)}\\
\hline
$(++)_{00}$ & $(-+)_{00}$ & $(+-)_{00}$ & $(--)_{00}$ \\
\hline
$1.8042 -i0.3971$  & $1.8013 -i0.2575$   & 
$1.6693 -i0.2178$   & $1.6528 -i0.2245$ \\
$1.8068 -i0.2664$ & $1.8325 -i0.4072$ & 
$1.7904 -i0.4035$ & $1.8189 -i0.4207$ \\
\hline
$1.8042 +i0.3971$  & $1.7648 +i0.5187$   & 
$1.7697 +i0.3874$ & $1.7178 +i0.4407$ \\
$1.8068 +i0.2664$ & $1.8251 +i0.2468$ & 
$1.8415 +i0.5646$ & $1.7793 +i0.6420$ \\
\hline
\end{tabular}
\end{center}
\caption{\sf
Poles of the two-channel $S$-matrix on all the principal sheets of the Riemann
surface within a distance of $\sim1$\,MeV from the central point, 
$E_0=1.8$\,MeV, of the expansions (\ref{Aapprox}, \ref{Bapprox}).
The energy is counted from the $p\,{}^7\mathrm{Be}$ threshold.
}
\label{table.allpoles}
\end{table}

The resonance energies obtained with different $E_0$, are statistically 
independent. We assume that they have the normal distribution and  calculate the 
corresponding average values as well as the standard deviations. The results of 
these calculations (statistical averaging) are given in 
Table~\ref{table.resonances.average}, where for the purpose of comparison, we 
also put the parameters of the $0^+$ resonance obtained in 
Ref.~\cite{Rogachev2013}.

By scanning all four principal sheets of the Riemann surface within a distance
of $\sim1$\,MeV around the central energy $E_0$, we found several $S$-matrix
poles on each of the sheets. These calculations were done with 
$E_0=1.8$\,MeV. Thus found poles are listed in Table~\ref{table.allpoles}.

\begin{figure}
\centerline{\epsfig{file=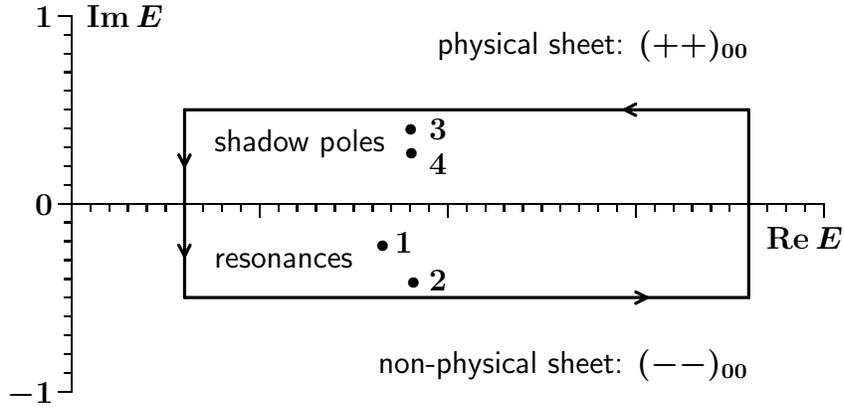}}
\caption{\sf
The $S$-matrix poles adjacent to the real axis of the scattering energies
where the physical and non-physical sheets of the Riemann surface are connected.
The corresponding energies and the $S$-matrix 
residues are given in Tables \ref{table.allpoles} and \ref{table.Residues}. The 
numerical labels of
the poles are used as the corresponding references in Fig.~\ref{fig.MLexcl}.
The rectangular contour is used for the integration in the Mittag-Leffler sum
(\ref{MittagLeffler}).
}
\label{fig.4poles}
\end{figure}

Among all the poles, only those that 
are adjacent to the physical scattering energies, may influence the physical 
observables. They are those given in the 
left bottom and right top blocks of Table~\ref{table.allpoles}. There are four 
of them: two resonances on the sheet $(--)_{00}$ and two poles on the physical 
sheet $(++)_{00}$. They are depicted in Fig.~\ref{fig.4poles}. 

The sheets 
$(--)_{00}$ and $(++)_{00}$ are cut along the real axis. At the energies 
above the second threshold, the upper rim of the $(--)_{00}$-cut is connected 
to the lower rim of the $(++)_{00}$-cut. The connecting line is the real axis 
of the physical scattering energies. Thanks to the connection, it is possible 
to continuously move from the sheet $(--)_{00}$ to $(++)_{00}$ and back, for 
example, along the rectangular contour shown in Fig.~\ref{fig.4poles}.

In contrast to the resonances, the solutions of the Schr\"odinger equation, 
corresponding to the complex poles on the sheet $(++)_{00}$, have an unphysical 
behaviour (in particular, they have unphysical time dependence). The physical 
system cannot be in such a state, but mathematically the poles exist anyway and 
may influence the behaviour of the $S$-matrix on the real axis. Such poles are 
sometimes called shadow ones. The influence of these poles on the scattering 
cross section is explored in the next section.

\subsection{Contributions from individual poles}
\label{sec.Mittag-Leffler}
The $S$-matrix has many poles on the Riemann surface. Even 
just on the principal sheets and only around the resonance 
energy ($E\sim1.8\,\mathrm{MeV}$), it has eight poles given in 
Table~\ref{table.allpoles}. Of course, their influence on the scattering cross 
sections are different. Apparently, the poles that are far away from the axis 
of the real scattering energies, contribute very little, if any. This axis 
passes through the connection of the sheets $(++)_{00}$ and $(--)_{00}$
(see Fig.~\ref{fig.sheets_2ch_Coulomb}). Therefore the only noticeable 
influence can be expected from the four poles shown in Fig.~\ref{fig.4poles}, 
which are near that axis.

It is always useful to know how important each 
individual pole is. A reasonable answer to such a question can be obtained by 
decomposing the $S$-matrix in a sum of the pole terms and the background 
integral. Such a decomposition is possible thanks to the Mittag-Leffler theorem 
known in the complex analysis. In fact, for our purpose, it is sufficient to 
apply a more simple residue theorem, which leads to the Mittag-Leffler 
decomposition (see Refs.~\cite{two_channel} and \cite{our_He5}).

Consider the rectangular contour shown in Fig.~\ref{fig.4poles},
which encloses the four chosen poles. If $E$ is a point inside this contour 
(for calculating the cross section, we choose it on the real axis), then
according to the residue theorem, we have
\begin{equation}
\label{Res_theorem}
   \oint\frac{\bm{S}(\zeta)}{\zeta-E}d\zeta=
   2\pi i\bm{S}(E)+2\pi i\sum_{j=1}^L
   \frac{\mathrm{Res}[\bm{S},E_j]}{E_j-E}\ ,
\end{equation}
where $E_j$ are the poles ($L=4$). This gives
\begin{equation}
\label{MittagLeffler}
   \bm{S}(E)=\sum_{j=1}^L
   \frac{\mathrm{Res}[\bm{S},E_j]}{E-E_j}+
   \frac{1}{2\pi i}\oint\frac{\bm{S}(\zeta)}{\zeta-E}d\zeta\ ,
\end{equation}
which is a particular form of the Mittag-Leffler decomposition, where
the matrix $\bm{S}(E)$ is written as a sum of the individual pole contributions 
and a background integral.

\begin{table}
\begin{center}
\begin{tabular}{|c|c|c|c|}
\hline
\sffamily sheet & \sffamily pole: $E$\,(MeV) &
\sffamily $\mathrm{Res}[S_{nn'},E]$\,(MeV) & $n,n'$ \\
\hline
\hline
\multirow{8}{*}{$(--)_{00}$} &
\multirow{4}{*}{$1.6528-i0.2245$} & 
             $-0.0138-i0.0072$   & 1,1 \\
&&           $\phantom{+}0.0433+i0.0261$ & 1,2 \\
&&           $\phantom{+}0.0328+i0.0261$    & 2,1 \\
&&           $-0.1014-i0.0918$   & 2,2 \\
\cline{2-4}		
&
\multirow{4}{*}{$1.8189-i0.4207$} & 
              $\phantom{+}0.0291-i0.0147$ &  1,1 \\
&           & $-0.0028-i0.0207$ &  1,2 \\
&           & $\phantom{+}0.0134-i0.0156$ &  2,1 \\
&           & $-0.0066-i0.0114$   &  2,2 \\
\hline
\hline
\multirow{8}{*}{$(++)_{00}$} &
\multirow{4}{*}{$1.8042+i0.3971$} & 
             $-0.0271-i0.0131$ & 1,1 \\
&          & $\phantom{+}0.0148-i0.0210$ & 1,2 \\
&          & $-0.0102-i0.0249$ & 2,1 \\
&          & $\phantom{+}0.0224-i0.0052$ & 2,2 \\
\cline{2-4}		
&
\multirow{4}{*}{$1.8068+i0.2664$} & 
              $\phantom{+}0.0060+i0.0088$ &  1,1 \\
&           & $-0.0221-i0.0441$ &  1,2 \\
&           & $-0.0096-i0.0288$ &  2,1 \\
&           & $\phantom{+}0.0266+i0.1381$ &  2,2 \\
\hline
\end{tabular}
%
%
%
\end{center}
\caption{\sf
Poles of the two-channel $S$-matrix and the corresponding residues of its
elements in the domains of the Riemann sheets $(--)_{00}$ and $(++)_{00}$
adjacent to the axis of the real scattering energies (see
Fig.~\ref{fig.4poles}). The energy is counted from the 
$p\,{}^7\mathrm{Be}$-threshold.
}
\label{table.Residues}
\end{table}

For any given scattering energy $E$, the background integral can be found by 
numerical integration of the $S$-matrix, which we obtained after fitting the 
experimental data. We assume that all the poles of the $S$-matrix 
(\ref{Smatrix}) are simple, i.e. 
\begin{equation}
\label{detFsimpleZero}
   \det\bm{f}^{(\rm in)}(E)
   \ \mathop{\longrightarrow}\limits_{E\to E_j}
   \ \mathrm{const}\cdot(E-E_j)\ .
\end{equation}
Therefore the residues of the $S$-matrix at the poles can be found by 
numerical differentiation of the determinant of
the Jost matrix,
\begin{equation}
\label{ResidueExplicit}
    {\rm Res}\,[\bm{S},E]=\bm{f}^{(\rm out)}(E)\left(
    \begin{array}{cc}
    f^{(\rm in)}_{22}(E) & -f^{(\rm in)}_{12}(E) \\[3mm]
    -f^{(\rm in)}_{21}(E) & f^{(\rm in)}_{11}(E)
    \end{array}\right)
    \left[\frac{d}{dE}\det \bm{f}^{(\rm in)}(E)\right]^{-1}\ .
\end{equation}
Thus calculated residues for the four poles are
given in Table \ref{table.Residues}. Using these residues and the
numerically calculated background integral, we obtained (as it should be) 
exactly the same cross sections that are shown in Figs.
\ref{fig.fit.11}, \ref{fig.fit.12}, \ref{fig.fit.21}, and \ref{fig.fit.22}.
This is a kind of cross-check of our calculations. 

\begin{figure}
\centerline{\epsfig{file=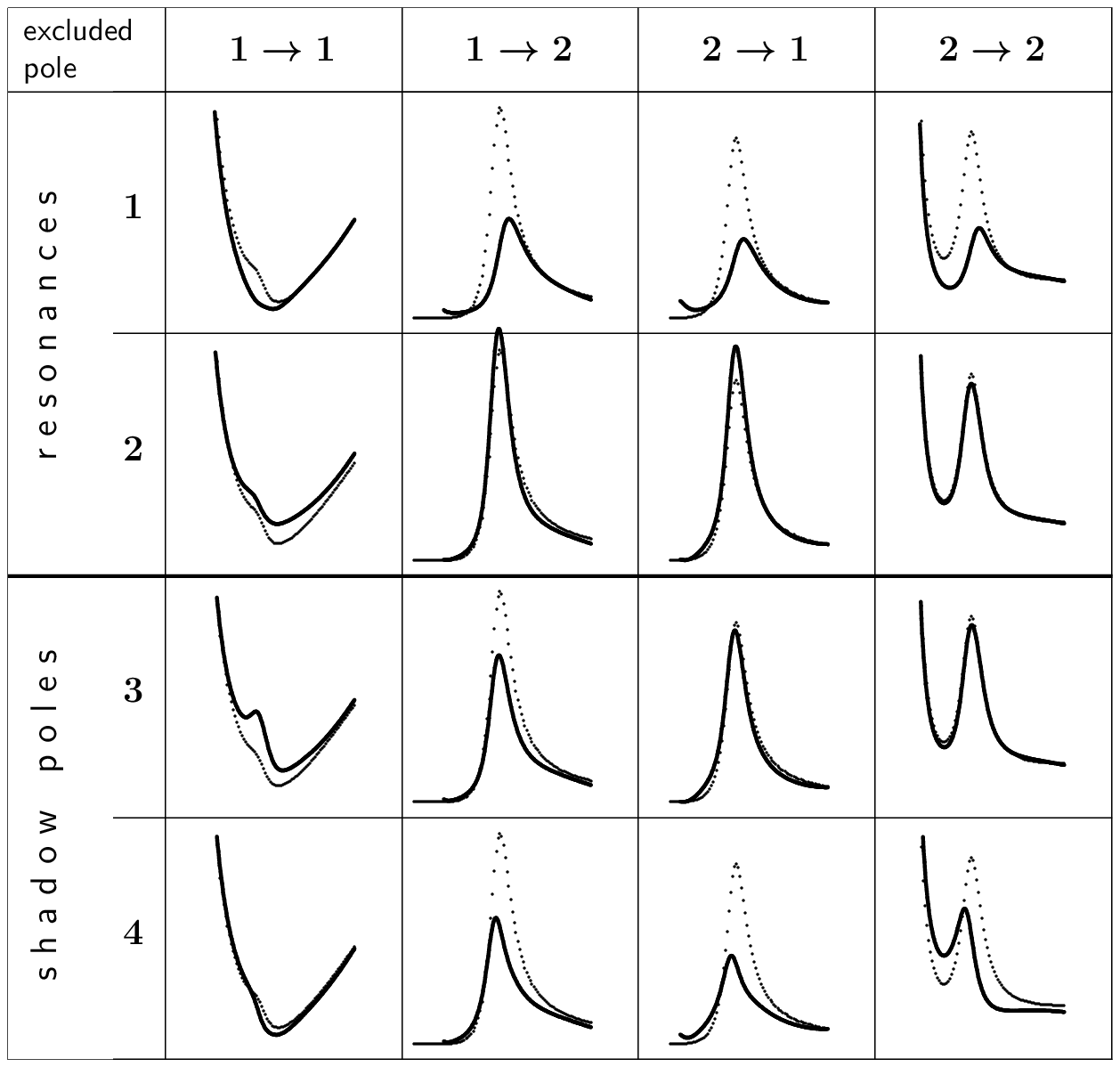}}
\caption{\sf
The dots represent the experimental (i.e. the $R$-matrix)
cross sections for the inter-channel transitions $n\to n'$, where  the
channels are labeled as in Eq.~(\ref{coupled_gammas}).
The curves show the corresponding cross sections obtained when
a single pole is excluded from the Mittag-Leffler sum (\ref{MittagLeffler}).  
}
\label{fig.MLexcl}
\end{figure}

Now, in order to get an idea of the role of each pole, we can omit them one by 
one from the sum (\ref{MittagLeffler}) and see how this affects the partial 
cross sections. The results of such pole exclusions are shown in 
Fig.~\ref{fig.MLexcl}. The curves show the cross sections when one pole is 
excluded. The dots are the experimental data (i.e. the $R$-matrix cross 
sections).

\begin{figure}
\centerline{\epsfig{file=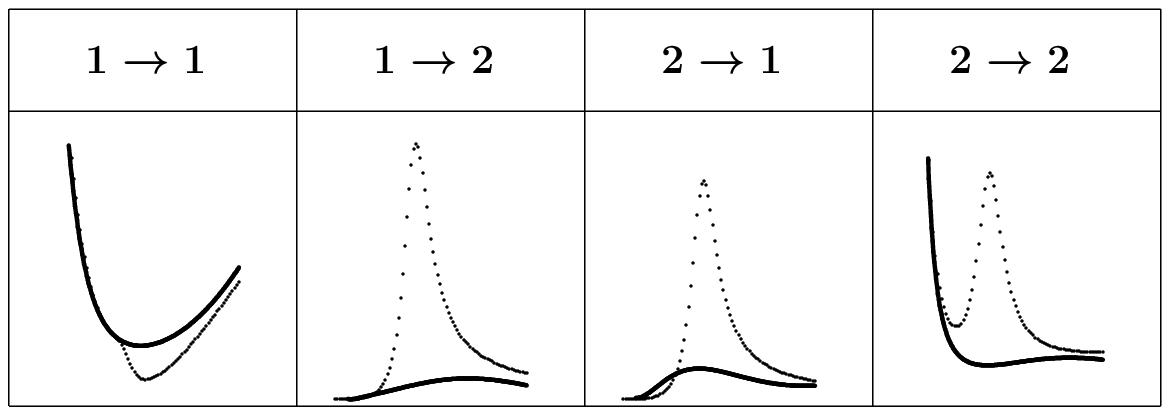}}
\caption{\sf
The background-scattering contributions (curves) to the partial cross sections 
for the inter-channel transitions $n\to n'$, when all
four poles shown in Fig.~\protect{\ref{fig.4poles}} are excluded from the
Mittag-Leffler sum (\ref{MittagLeffler}). The dots are the corresponding
experimental (i.e. the $R$-matrix) cross sections.
}
\label{fig.Nopoles}
\end{figure}

It is seen that the second resonance (pole number 2) contributes very little 
and mainly to the elastic cross section in the first channel. The influences of 
the other three poles are noticeable in various channels.
It is also interesting to know what happens if we only leave the
background integral and exlude all the pole terms
from the Mittag-Leffler expansion (\ref{MittagLeffler}). The result of such an
exclusion can be seen in Fig.~\ref{fig.Nopoles}. The background term 
describes the general behaviour of the cross sections and gives a 
reasonable approximation for them to the left and to the right of the resonance 
energies. However, inside the resonance energy-interval, without the resonant 
and the shadow poles all the cross sections are far from the experimental 
points. 

\section{Summary and conclusion}
\label{sec.conclusion}
As was stated in the Introduction, the main task of the present work was to 
confirm the existence and to accurately determine the
parameters of the lowest resonance level with the quantum numbers $J^\pi=0^+$ 
in the spectrum of the eight-nucleon system ${}^8\mathrm{B}$.
For this purpose, we constructed the two-channel Jost-matrices
that have proper analytic structure and are defined on the Riemann surface with 
the proper topology (with both the square-root and logarithmic branching). The 
free parameters of these Jost matrices were fixed using an available $R$-matrix 
fit\cite{Rogachev2013} of experimental data on  $p\,{}^7\mathrm{Be}$ scattering.

Exploring the behaviour of these Jost matrices on the principal sheets of the 
Riemann surface, we located 16 poles of the $S$-matrix (see 
Table~\ref{table.allpoles}). Among them, only four poles (two resonances and two 
shadow poles) are located close enough to the axis of the real scattering 
energies and therefore can influence the observable cross sections (see 
Fig.~\ref{fig.4poles}). 

Therefore, we found that instead of a single $0^+$ resonance, there are two 
overlapping resonances with almost the same parameters as were reported in 
Ref.~\cite{Rogachev2013} (see Table~\ref{table.resonances.average}). In 
addition to them, there are also two overlapping shadow poles on the opposite 
side of the real axis.

In order to isolate the individual contributions to the $S$-matrix from the 
resonances and the shadow poles, we used the Mittag-Leffler decomposition.
In this way it was established that the second resonance has a rather weak 
influence on the energy dependencies of the partial cross sections. The roles 
of the other three poles are noticeable.

As is seen from Fig.~\ref{fig.MLexcl}, the first resonance and the second 
shadow pole significantly change the inelastic cross sections and the
elastic scattering in the second channel. In principle, such changes could be 
detected experimentally, if the ${}^7\mathrm{Be}$ target is exposed to 
$\gamma$-rays of the energy $\sim0.5\,\mathrm{MeV}$, when the cross section 
of $p\,{}^7\mathrm{Be}$ collision is being measured. In such a case, the 
electromagnetic radiation could cause part of the target nuclei to transit from 
the ground to the first excited state,
${}^7\mathrm{Be}(\frac32^-)+\gamma\to{}^7\mathrm{Be}^*(\frac12^-)$.


\end{document}